\documentclass[conference]{IEEEtran}
\IEEEoverridecommandlockouts
\usepackage{cite}
\usepackage{mathptmx}
\usepackage{amsmath,amssymb,amsfonts}
\usepackage{algorithmic}
\usepackage{graphicx}
\usepackage{textcomp}
\usepackage{xcolor}
\usepackage{booktabs}
\usepackage{multirow} 
\def\BibTeX{{\rm B\kern-.05em{\sc i\kern-.025em b}\kern-.08em
    T\kern-.1667em\lower.7ex\hbox{E}\kern-.125emX}}
\begin{document}

\title{EEG-based Sleep Staging with Hybrid Attention\\
\thanks{{$^{*}$Corresponding Author: Yang Liu and Jiaping Xiao.}
\\
{$^{1}$Xinliang Zhou, and Chenyu Liu are with the School of Computer Science and Engineering, Nanyang Technological University, 639798, Singapore {\tt\small\{xinliang001, chenyu003\}@e.ntu.edu.sg}}%
\\
{$^{3,1}$Yang Liu is with Zhejiang Sci-Tech University, Hangzhou, 314423, China and the School of Computer Science and Engineering, 50 Nanyang Avenue, 639798, Singapore
        {\tt\small \ {yangliu}@ntu.edu.sg}}%
\\
{$^{2}$Jiaping Xiao is with the School of Mechanical and Aerospace Engineering, Nanyang Technological University, 639798, Singapore
        {\tt\small \ {jiaping001}@e.ntu.edu.sg}}%
}}

 \author{Xinliang Zhou$^1$\hspace{1cm}Chenyu Liu$^1$\hspace{1cm}$^{*}$Jiaping Xiao$^2$ \hspace{1cm}$^{*}$Yang Liu$^3{}^,{}^1$\\

$^1$School of Computer Science and Engineering, Nanyang Technological University\\$^2$School of Mechanical and Aerospace Engineering, Nanyang Technological University\\
$^3$Zhejiang Sci-Tech University\\

\{xinliang001, chenyu003, jiaping001\}@e.ntu.edu.sg, 
\{yangliu\}@ntu.edu.sg
}




\maketitle
\begin{abstract}
Sleep staging is critical for assessing sleep quality and diagnosing sleep disorders. However, capturing both the spatial and temporal relationships within electroencephalogram (EEG) signals during different sleep stages remains challenging. In this paper, we propose a novel framework called the Hybrid Attention EEG Sleep Staging (HASS) Framework. Specifically, we propose a well-designed spatio-temporal attention mechanism to adaptively assign weights to inter-channels and intra-channel EEG segments based on the spatio-temporal relationship of the brain during different sleep stages. Experiment results on the MASS and ISRUC datasets demonstrate that HASS can significantly improve typical sleep staging networks. Our proposed framework alleviates the difficulties of capturing the spatial-temporal relationship of EEG signals during sleep staging and holds promise for improving the accuracy and reliability of sleep assessment in both clinical and research settings.
\end{abstract}


\section{Introduction}

Sleep staging is a crucial process in evaluating sleep quality and diagnosing sleep disorders, which involves dividing a sleep period into several periodical stages \cite{zhou2023interpretable,liu2023bstt}. However, manual sleep staging is time-consuming, subjective, and requires professional expertise, which can lead to unstable and unreliable further sleep disorder diagnosis. Therefore, automated sleep staging methods, including deep learning-based approaches, have been developed to improve efficiency and accuracy.

Despite the progress in automated sleep staging, accurately capturing the spatio-temporal relationships within EEG signals during different sleep stages remains challenging. Previous studies have proposed several methods, such as earlier traditional machine learning methods and current deep learning methods, to enhance the performance of sleep staging systems. However, these methods mainly consider extracting only spatial or temporal features inside the EEG signals, and they have limitations in capturing the complex spatio-temporal relationships of the brain.

To address the above challenges, this paper proposes a novel hybrid attention EEG sleep staging (HASS) framework. The framework employs a well-designed encoder based on the attention mechanism that adaptively assigns weights to different EEG segments and channels based on their spatio-temporal relationships during sleep stages. Specifically, the proposed encoder internally contains two components based on the attention mechanism: intra-channel and inter-channel attention. The two components are responsible for capturing the spatial and temporal relationships in sleep EEG signals, respectively. The captured spatial and temporal relationships are further integrated as the spatio-temporal relationships, which can effectively improve sleep staging networks' performance.

The proposed HASS framework has shown promising results in improving the F1 score and accuracy of typical sleep staging networks, as demonstrated by experimental results on the MASS \cite{o2014montreal} and ISRUC \cite{khalighi2016isruc} datasets. By capturing the complex spatio-temporal relationships of EEG signals during sleep staging, the HASS framework shows excellent potential for improving the stability and reliability of sleep assessment in both clinical and research settings.

\begin{figure*}[ht] 
\centering 
\includegraphics[width=1.7 \columnwidth]{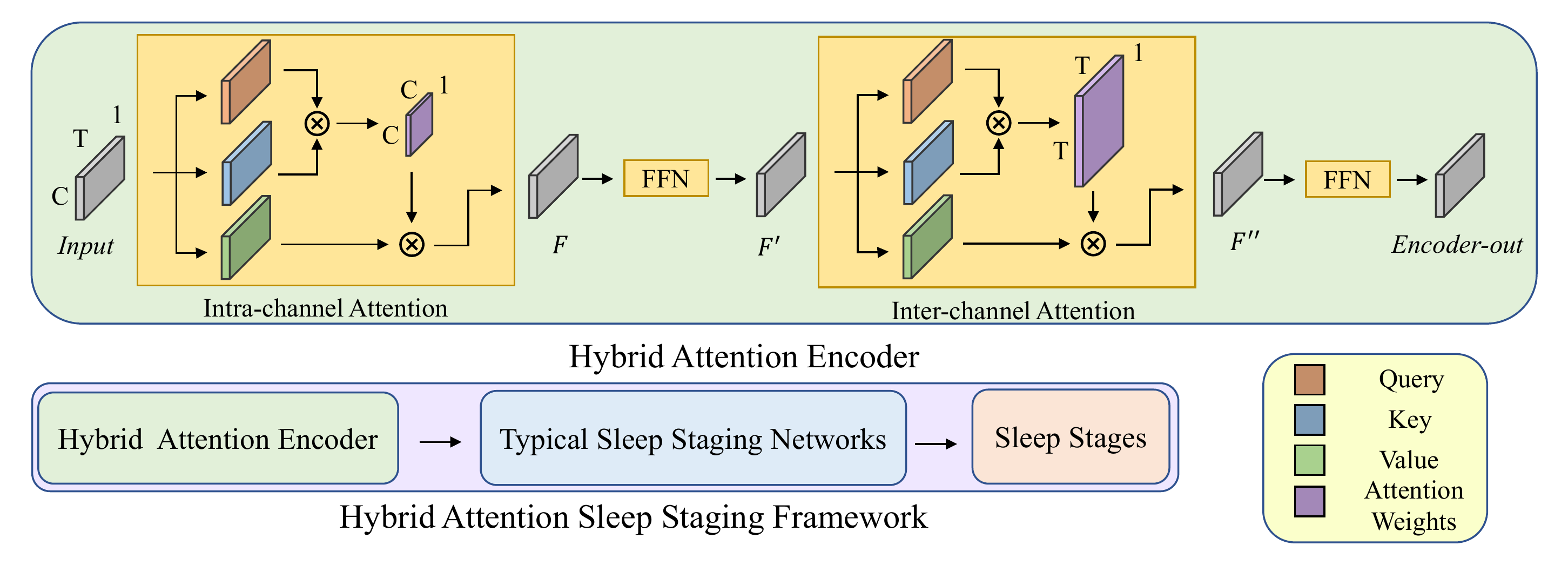}
\caption{Overall of the Hybrid Attention Sleep Staging Framework} 
\label{} 
\end{figure*}

\section{Related Work}
Recognizing different sleep stages is crucial for diagnosing and treating sleep disorders. In the past, support vector machines (SVM) \cite{alickovic2018ensemble} and random forests (RF) \cite{fraiwan2012automated} were widely used for sleep staging. However, they require extensive prior knowledge and manual feature engineering as well as suffer limited performance. Nowadays, deep learning approaches are the primary method for sleep staging and illustrate better performance. 

In the early stage, deep learning methods, such as convolutional neural networks (CNN), are utilized to extract temporal features from sleep signals \cite{chambon2018deep}. For instance, Dong \emph{et al.} \cite{chambon2018deep} proposed the multivariate CNN to capture the temporal features for sleep staging. After that, the recurrent neural networks (RNN) \cite{phan2018dnn,phan2019seqsleepnet}, the combination of RNN and CNN \cite{supratak2017deepsleepnet}, and spiking neural networks (SNN) \cite{jia2022hybrid} are also applied to extract the temporal representation in different sleep stages. 
Apart from mining the temporal features to achieve sleep stages classification, some studies have tried to extract the spatial features in the sleep data. Liu \emph{et al.}  \cite{liubstt} set electrodes as nodes and mine the spatial features between different channels using relational thinking networks (RTN). 

The aforementioned methods only extract temporal and spatial features in sleep data separately, which ignores the spatio-temporal correlation between features. In order to better utilize the spatio-temporal relationships in sleep data, this paper proposes a HASS framework to enhance the typical sleep staging networks' performance.



\section{Methodology}

\subsection{Description of Matrix Q, K and V}
The Query (Q), Key (K), and Value (V) matrices play an essential role in the self-attention mechanism, which is designed to capture long-range dependencies and complex relationships in input time sequences. These matrices are derived from the input representations through linear transformations using three different weight matrices, namely $W_Q$, $W_K$, and $W_V$.

The Query (Q) matrix represents the transformed state of the current input element, enabling the model to search for relevant context within the input sequence. In essence, it serves as a basis for comparison against other input elements to determine their relevance. The Key (K) matrix encompasses the transformed context representations of the other input elements, allowing for comparing the Query and each context element. This comparison is crucial for calculating attention scores, which indicate the relative importance of each element in the time sequence. The Value (V) matrix retains the original input element representations used to compute the final attention-weighted output. This output serves as a context-aware representation of the current input element, considering the relationships between the element and its surroundings.
\subsection{Hybrid Attention Framework}
\begin{table*}[t]
\centering
\caption{The comparison of the performance (F1 Score and Accuracy) of MASS and ISRUC dataset between the original four sleep staging networks and them after applying HASS. }
\label{result}
\scalebox{0.8}{
\begin{tabular}{@{}cccccllllccccllc@{}}
\toprule
\toprule
\multicolumn{1}{l}{} &
  \multicolumn{1}{l}{} &
  \multicolumn{7}{c}{MASS Dataset} &
  \multicolumn{7}{c}{ISRUC Dataset} \\ \midrule
\multicolumn{1}{l}{\multirow{2}{*}{Typical Network}} &
  \multirow{2}{*}{HASS or Not} &
  \multicolumn{2}{c}{Overall Results} &
  \multicolumn{5}{c}{F1 Score for Each Stages} &
  \multicolumn{2}{c}{Overall Results} &
  \multicolumn{5}{c}{F1 Score for Each Stages} \\ \cmidrule(l){3-16} 
\multicolumn{1}{l}{} &
   &
  \multicolumn{1}{l}{F1 Score} &
  \multicolumn{1}{l}{Accuracy} &
  W &
  \multicolumn{1}{c}{N1} &
  \multicolumn{1}{c}{N2} &
  \multicolumn{1}{c}{N3} &
  \multicolumn{1}{c}{REM} &
  F1 Score &
  Accuracy &
  W &
  N1 &
  \multicolumn{1}{c}{N2} &
  \multicolumn{1}{c}{N3} &
  REM \\ \midrule
\textbf{TSN \cite{supratak2020tinysleepnet}} &
  \begin{tabular}[c]{@{}c@{}}Yes\\ No\end{tabular} &
  \begin{tabular}[c]{@{}c@{}}\textbf{0.811}\\ 0.798\end{tabular} &
  \begin{tabular}[c]{@{}c@{}}\textbf{0.881}\\ 0.858\end{tabular} &
  \begin{tabular}[c]{@{}c@{}}\textbf{0.885}\\ 0.873\end{tabular} &
  \begin{tabular}[c]{@{}l@{}}\textbf{0.565}\\ 0.547\end{tabular} &
  \begin{tabular}[c]{@{}l@{}}\textbf{0.892}\\ 0.888\end{tabular} &
  \begin{tabular}[c]{@{}l@{}}0.847\\ \textbf{0.848}\end{tabular} &
  \begin{tabular}[c]{@{}l@{}}\textbf{0.921}\\ 0.889\end{tabular} &
  \begin{tabular}[c]{@{}c@{}}\textbf{0.774}\\ 0.751\end{tabular} &
  \begin{tabular}[c]{@{}c@{}}\textbf{0.794}\\ 0.769\end{tabular} &
  \begin{tabular}[c]{@{}c@{}}\textbf{0.891}\\ 0.868\end{tabular} &
  \begin{tabular}[c]{@{}c@{}}\textbf{0.576}\\ 0.545\end{tabular} &
  \begin{tabular}[c]{@{}l@{}}\textbf{0.874}\\ 0.855\end{tabular} &
  \begin{tabular}[c]{@{}l@{}}\textbf{0.884}\\ 0.878\end{tabular} &
  \begin{tabular}[c]{@{}c@{}}\textbf{0.847}\\ 0.828\end{tabular} \\ \cmidrule(l){2-16} 
\textbf{DSN \cite{supratak2017deepsleepnet}} &
  \begin{tabular}[c]{@{}c@{}}Yes\\ No\end{tabular} &
  \begin{tabular}[c]{@{}c@{}}\textbf{0.791}\\ 0.773\end{tabular} &
  \begin{tabular}[c]{@{}c@{}}\textbf{0.855}\\ 0.834\end{tabular} &
  \begin{tabular}[c]{@{}c@{}}\textbf{0.873}\\ 0.861\end{tabular} &
  \begin{tabular}[c]{@{}l@{}}\textbf{0.528}\\ 0.514\end{tabular} &
  \begin{tabular}[c]{@{}l@{}}\textbf{0.888}\\ 0.861\end{tabular} &
  \begin{tabular}[c]{@{}l@{}}0.821\\ \textbf{0.825}\end{tabular} &
  \begin{tabular}[c]{@{}l@{}}\textbf{0.883}\\ 0.864\end{tabular} &
  \begin{tabular}[c]{@{}c@{}}\textbf{0.755}\\ 0.731\end{tabular} &
  \begin{tabular}[c]{@{}c@{}}\textbf{0.781}\\ 0.756\end{tabular} &
  \begin{tabular}[c]{@{}c@{}}\textbf{0.883}\\ 0.852\end{tabular} &
  \begin{tabular}[c]{@{}c@{}}\textbf{0.535}\\ 0.527\end{tabular} &
  \begin{tabular}[c]{@{}l@{}}\textbf{0.861}\\ 0.829\end{tabular} &
  \begin{tabular}[c]{@{}l@{}}\textbf{0.889}\\ 0.879\end{tabular} &
  \begin{tabular}[c]{@{}c@{}}\textbf{0.831}\\ 0.796\end{tabular} \\ \cmidrule(l){2-16} 
\textbf{MCNN \cite{chambon2018deep}} &
  \begin{tabular}[c]{@{}c@{}}Yes\\ No\end{tabular} &
  \begin{tabular}[c]{@{}c@{}}\textbf{0.818}\\ 0.804\end{tabular} &
  \begin{tabular}[c]{@{}c@{}}\textbf{0.875}\\ 0.862\end{tabular} &
  \begin{tabular}[c]{@{}c@{}}\textbf{0.895}\\ 0.884\end{tabular} &
  \begin{tabular}[c]{@{}l@{}}\textbf{0.553}\\ 0.548\end{tabular} &
  \begin{tabular}[c]{@{}l@{}}\textbf{0.918}\\ 0.898\end{tabular} &
  \begin{tabular}[c]{@{}l@{}}\textbf{0.836}\\ 0.825\end{tabular} &
  \begin{tabular}[c]{@{}l@{}}\textbf{0.917}\\ 0.897\end{tabular} &
  \begin{tabular}[c]{@{}c@{}}\textbf{0.782}\\ 0.764\end{tabular} &
  \begin{tabular}[c]{@{}c@{}}\textbf{0.801}\\ 0.782\end{tabular} &
  \begin{tabular}[c]{@{}c@{}}\textbf{0.902}\\ 0.885\end{tabular} &
  \begin{tabular}[c]{@{}c@{}}\textbf{0.603}\\ 0.571\end{tabular} &
  \begin{tabular}[c]{@{}l@{}}\textbf{0.861}\\ 0.851\end{tabular} &
  \begin{tabular}[c]{@{}l@{}}0.893\\ \textbf{0.894}\end{tabular} &
  \begin{tabular}[c]{@{}c@{}}\textbf{0.842}\\ 0.833\end{tabular} \\ \cmidrule(l){2-16} 
\textbf{U-Time \cite{perslev2019u}} &
  \begin{tabular}[c]{@{}c@{}}Yes\\ No\end{tabular} &
  \begin{tabular}[c]{@{}c@{}}\textbf{0.794}\\ 0.782\end{tabular} &
  \begin{tabular}[c]{@{}c@{}}\textbf{0.873}\\ 0.854\end{tabular} &
  \begin{tabular}[c]{@{}c@{}}\textbf{0.885}\\ 0.865\end{tabular} &
  \begin{tabular}[c]{@{}l@{}}\textbf{0.528}\\ 0.517\end{tabular} &
  \begin{tabular}[c]{@{}l@{}}\textbf{0.891}\\ 0.882\end{tabular} &
  \begin{tabular}[c]{@{}l@{}}\textbf{0.813}\\ 0.801\end{tabular} &
  \begin{tabular}[c]{@{}l@{}}\textbf{0.883}\\ 0.875\end{tabular} &
  \begin{tabular}[c]{@{}c@{}}\textbf{0.727}\\ 0.713\end{tabular} &
  \begin{tabular}[c]{@{}c@{}}\textbf{0.769}\\ 0.755\end{tabular} &
  \begin{tabular}[c]{@{}c@{}}\textbf{0.869}\\ 0.844\end{tabular} &
  \begin{tabular}[c]{@{}c@{}}0.524\\ \textbf{0.525}\end{tabular} &
  \begin{tabular}[c]{@{}l@{}}\textbf{0.813}\\ 0.793\end{tabular} &
  \begin{tabular}[c]{@{}l@{}}\textbf{0.885}\\ 0.861\end{tabular} &
  \begin{tabular}[c]{@{}c@{}}\textbf{0.769}\\ 0.755\end{tabular} \\ \bottomrule
  \bottomrule
\end{tabular}}
\end{table*} 

 The hybrid attention framework contains two parts. The first is the novel hybrid attention encoder, and the second is the typical sleep staging network. The hybrid attention encoder captures the spatial and temporal relationships in the EEG signals, and the typical sleep staging network is utilized to achieve the classification. Specially inside the hybrid attention encoder, the intra-channel attention, inter-channel attention, and FFN model are described respectively in the following. \\

\subsubsection{Intra-channel Attention} The intra-channel attention encoder can capture the spatial relationship between the channels. We set $Input$ as $I$, where
$I\in \mathbb{R}^{C \times T \times 1}$, the encoder process the $Input$ into $F^{\prime} \in \mathbb{R}^{C \times T \times 1  }$ as follows:
{\footnotesize
\begin{equation}
\begin{aligned}
F & =\operatorname{LN}\left(I+\operatorname{DA}\left(I_{\text {Q }}, I_{\text {K}}, I_{\text {V }}\right) ; \Theta, \boldsymbol{\Phi}\right), \\
\end{aligned}
\end{equation}
\begin{equation}
\begin{aligned}
F^{\prime} & =\operatorname{LN}\left(F+\operatorname{FFN}\left(F ; \boldsymbol{\Psi}\right)\right).
\end{aligned}
\end{equation}
}

\subsubsection{Inter-channel Attention}
Unlike the intra-channel attention encoder, the inter-channel one makes the DA inside each channel, which can capture the temporal relationship from the sleep signals. The calculations are conducted as follows:
{\footnotesize
\begin{equation}
\begin{aligned}
F^{\prime \prime} & =\operatorname{LN}\left(F^{\prime}+\operatorname{DA}\left(F^{\prime}_{\text {Q}}, F^{\prime}_{\text {K}}, F^{\prime}_{\text {V}}\right) ; \Theta, \boldsymbol{\Phi}\right), \\
\end{aligned}
\end{equation}
\begin{equation}
\begin{aligned}
Encoder\_out & =\operatorname{LN}\left(F^{\prime \prime}+\operatorname{FFN}\left(F^{\prime \prime} ; \boldsymbol{\Psi}\right)\right),
\end{aligned}
\end{equation}
}
where LN denotes the layer normalization \cite{ba2016layer} and DA denotes dot-product attention. To be specific, given query $Q \in \mathbb{R}^{d_k \times N}$, key $K \in \mathbb{R}^{d_k \times N}$, and $d_v$-dimensional value $V \in \mathbb{R}^{d_v \times N}$ inputs, $\mathrm{DA}$ is calculated as:
{\footnotesize

\begin{equation}
\begin{aligned}
& Q^{(n)}=W_Q^{(n)} Q+\boldsymbol{b}_Q^{(n)} \mathbf{1}^{\top} \in \mathbb{R}^{\frac{d_k}{m} \times N}, \\
\end{aligned}
\end{equation}
\begin{equation}
\begin{aligned}
& K^{(n)}=W_K^{(n)} K+\boldsymbol{b}_K^{(n)} \mathbf{1}^{\top} \in \mathbb{R}^{\frac{d_k}{m} \times N}, \\
\end{aligned}
\end{equation}
\begin{equation}
V^{(n)}=W_V^{(n)} V+\boldsymbol{b}_V^{(n)} \mathbf{1}^{\top} \in \mathbb{R}^{\frac{d_v}{m} \times N},
\end{equation}
\begin{equation}
\begin{aligned}
& \operatorname{DA}(Q, K, V ; \boldsymbol{\Theta}, \boldsymbol{\Phi})=W_O\left[\begin{array}{c}
V^{(1)} A^{(1) \top} \\
\vdots \\
V^{(m)} A^{(m) \top}
\end{array}\right]+\boldsymbol{b}_O \mathbf{1}^{\top} \in \mathbb{R}^{d_v \times N}, \\
\end{aligned}
\end{equation}
\begin{equation}
\begin{aligned}
& A^{(n)}=\operatorname{softmax}\left(\frac{Q^{(n) \top} K^{(n)}}{\sqrt{d_k / m}}\right) \in(0,1)^{N \times N}, \\
\end{aligned}
\end{equation}
}
where $m$ is the number of heads, $n \in\{1, \ldots, m\}$ is the index the head. The set of parameters $\Theta$ and $\Phi$ are defined as:
{\footnotesize
\begin{equation}
\begin{aligned}
& \boldsymbol{\Theta}:=\bigcup_{1 \leq i \leq m}\left\{W_Q^{(n)}, \boldsymbol{b}_Q^{(n)}, W_K^{(n)}, \boldsymbol{b}_K^{(n)}\right\}, \\
\end{aligned}
\end{equation}
\begin{equation}
\begin{aligned}
& \boldsymbol{\Phi}:=\left\{W_O, \boldsymbol{b}_O\right\} \cup \bigcup_{1 \leq i \leq m}\left\{W_V^{(n)}, \boldsymbol{b}_V^{(n)}\right\} .
\end{aligned}
\end{equation}
}

\subsection{Feed-forward Network (FFN)}
There are two FFN models in the hybrid attention sleep staging framework. Each FFN model consists of two dense layers and can be calculated as:
{\footnotesize
\begin{equation}
\begin{aligned}
\operatorname{FFN_1}\left(F ; \boldsymbol{\Psi}\right) & =\left(W_2\left[W_1 F+\boldsymbol{b}_1 \mathbf{1}^{\top}\right]_{+}+\boldsymbol{b}_2 \mathbf{1}^{\top}\right), \\
\boldsymbol{\Psi} & :=\left\{W_1, \boldsymbol{b}_1, W_2, \boldsymbol{b}_2\right\}.
\end{aligned}
\end{equation}

\begin{equation}
\begin{aligned}
\operatorname{FFN_2}\left(F^{\prime} ; \boldsymbol{\Psi}\right) & =\left(W_4\left[W_3 F^{\prime}+\boldsymbol{b}_3 \mathbf{1}^{\top}\right]_{+}+\boldsymbol{b}_4 \mathbf{1}^{\top}\right), \\
\boldsymbol{\Psi} & :=\left\{W_3, \boldsymbol{b}_3, W_4, \boldsymbol{b}_4\right\}.
\end{aligned}
\end{equation}
}
where $W_1, W_3 \in \mathbb{R}^{d_f \times D}$ and $W_2, W_4 \in \mathbb{R}^{D \times d_f}$ are mapping matrices, $\boldsymbol{b}_1, \boldsymbol{b}_3 \in \mathbb{R}^{d_f}$ and $\boldsymbol{b}_2, \boldsymbol{b}_4 \in \mathbb{R}^D$ are biases, and $[\cdot]_{+}$is the unit slope function.

\section{Experiments and Results}

\subsection{Dataset}
To thoroughly evaluate the efficacy of the HASS framework, we conduct experiments using two well-established datasets: the Institute of Systems and Robotics of the University of Coimbra (ISRUC) and the Montreal Sleep Study Archive-SS3 (MASS) datasets. The ISRUC dataset comprises PSG recordings from $100$ adult subjects, each with $6$ EEG channels. Meanwhile, the MASS dataset includes PSG records from $62$ adult subjects, each with $20$ EEG channels.
To standardize the data, we divide the recordings into time slices corresponding to sleep epochs, each representing $30$ seconds of sleep. To ensure the accuracy of our evaluations, we enlist the expertise of sleep specialists who manually classify each time slice into one of five different sleep stages: Wake (W), N1, N2, N3, and REM, following the AASM criteria \cite{berry2012aasm}.

\subsection{Settings}
In our experiment, the input is denoted as $I \in \mathbb{R}^{C \times T \times 1}$. Specifically, $C$ denotes the number of EEG channels, with $6$ channels in the ISRUC dataset and $20$ channels in the MASS dataset. $T$ signifies the time slices corresponding to sleep epochs, each denoting 30 seconds of sleep. Regarding intra-channel attention, the DA is calculated on channel dimension. Conversely, the DA is computed along the time slices dimension for inter-channel attention.

\subsection{Result and Analysis}
Table \ref{result} presents the comparison of the performance of four typical sleep staging networks, namely TinySleepNet (TSN) \cite{supratak2020tinysleepnet}, DeepSleepNet (DSN) \cite{supratak2017deepsleepnet}, MCNN \cite{chambon2018deep}, and U-Time \cite{perslev2019u}, on the MASS and ISRUC datasets, before and after applying the proposed HASS framework. The performance of the networks is evaluated using overall F1 scores, accuracies, and F1 scores for each sleep stage.

The results demonstrate that HASS can significantly improve the performance of the typical sleep staging networks. Specifically, for the MASS dataset, all four networks achieved higher overall F1 scores and accuracies when using HASS. The improvement in F1 scores for each stage is consistent across all networks. After applying HASS, the performance of the networks for all stages is improved. Notably, the most significant improvement was observed for the W stage, with all four networks achieving significantly higher F1 scores with HASS over the original networks. For instance, TSN and DSN achieved F1 scores of $0.885$ and $0.873$, respectively, with HASS, compared to $0.873$ and $0.861$ without HASS. Similarly, MCNN and U-Time achieved F1 scores of $0.895$ and $0.885$, respectively, with HASS, compared to $0.884$ and $0.865$ without HASS.

Likewise, for the ISRUC dataset, HASS improved the performance of all four networks, with higher overall F1 scores and accuracies over the original networks. However, the improvement in F1 scores is less significant compared to the MASS dataset, and three out of four networks achieved higher F1 scores for the W stage with HASS. Nevertheless, the improvement in F1 scores for all stages is consistent across all networks.

 The overall F1 scores and accuracy improvement are significant and consistent across all networks and for all sleep stages. These results suggest that the HASS framework can be highly effective in enhancing the accuracy of sleep staging networks, which is crucial for accurately diagnosing and treating sleep disorders. With these improvements, the HASS framework has the potential to significantly enhance the quality of sleep staging in clinical settings, thereby improving the overall quality of patient diagnosis and care.


\section{Conclusion}

The proposed HASS framework effectively improves the performance of typical sleep staging networks on two different datasets: MASS and ISRUC. The experiment results demonstrate that HASS can significantly enhance the overall F1 scores and accuracies of the networks and improve the performance of the networks for all sleep stages. Specifically, the most remarkable improvement was observed for the W stage on the MASS dataset. These findings suggest that the HASS method has the potential to enhance the accuracy of sleep staging networks, which is critical for the accurate diagnosis and treatment of sleep disorders. 

\section{ACKNOWLEDGMENTS}
This research is supported by National Research Foundation (NRF) Singapore, NRF Investigatorship NRF-NRFI06-2020-0001.

\bibliographystyle{IEEEbib}
\bibliography{ref}

\end{document}